\documentclass[aps,prl,twocolumn,showpacs,superscriptaddress,bibnotes]{revtex4}

\usepackage{graphicx}
\usepackage{dcolumn}
\usepackage{bm}

\begin{document}

\preprint{submitted to PRL}

\title{Driving mechanisms behind the various 
phases in manganese perovskites revealed by 
Mn $2p$ resonance photoemission}


\author{A. Sekiyama}
\email[]{sekiyama@mp.es.osaka-u.ac.jp}
\author{H. Fujiwara}
\author{A. Higashiya}
\author{S. Imada}
\affiliation{Department of Material Physics, 
Graduate School of Engineering Science, Osaka 
University, Toyonaka, Osaka 560-8531, Japan}
\author{H. Kuwahara}
\affiliation{Department of Physics, Sophia University, 
Tokyo 102-0094, Japan}
\author{Y. Tokura}
\affiliation{Department of Applied Physics, University of Tokyo, 
Tokyo 113-8656, Japan}
\author{S. Suga}
\affiliation{Department of Material Physics, 
Graduate School of Engineering Science, Osaka 
University, Toyonaka, Osaka 560-8531, Japan}


\date{\today}

\begin{abstract}
Unusual temperature and doping dependence of the Mn $3d$ spectral 
functions of manganese perovskites Nd$_{1-x}$Sr$_x$MnO$_3$ has been 
revealed by the high-resolution Mn $2p-3d$ resonance photoemission. 
The temperature-dependent spectra cannot be explained 
by any theoretical model currently under debate, while showing 
evidence for a microscopic and dynamic phase segregation. 
The experimental results strongly suggest that the competition of 
both the dynamical and static Jahn-Teller effects 
with ferromagnetic ordering at high and low temperatures, 
respectively, is responsible for the actual electronic states. 
\end{abstract}


\maketitle

Manganese perovskites have extensively been studied regarding 
colossal magnetoresistance (CMR)~\cite{CMRH,CMRT,Uru} and 
charge-ordering~\cite{KuwaSci} 
from both scientific and engineering viewpoints. 
Their electronic states have fundamentally been understood 
by the so-called double-exchange (DE) mechanism~\cite{ZenerDE,AHDE} 
for a long time. 
In the last decade, it has been recognized that a dynamical 
Jahn-Teller (JT) effect induced by a local lattice distortion also 
plays a substantial role.~\cite{MillisPRL96,MillisPRB96} 
Nowadays there are many theoretical models and calculations 
for the manganites, which well reproduce the temperature dependence of 
the resistivity~\cite{CMRT,Uru} and the optical 
conductivity.~\cite{Kaplan,Okimoto97,MWKim} 
Meanwhile the Mn $3d$ spectral functions and their temperature 
dependence obtained from angle-integrated photoemission 
are thought to be crucial for the test of these models, namely, to see 
whether these models really explain the Mn $3d$ spectral functions. 

\begin{figure}
\includegraphics[width=8.5cm,clip]{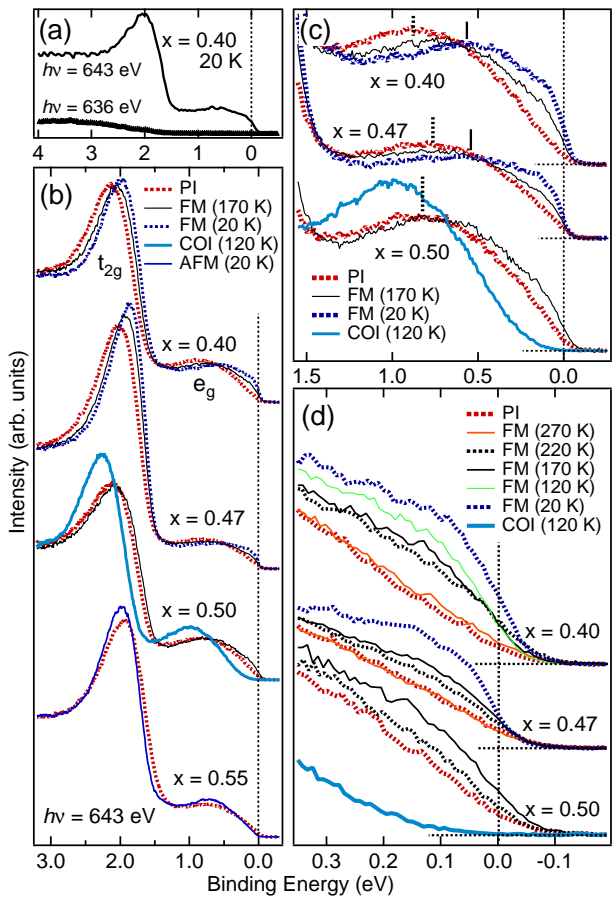}%
\caption{\label{Fig1}Mn $2p-3d$ resonance photoemission spectra of 
Nd$_{1-x}$Sr$_x$MnO$_3$ ($x$ = 0.40, 0.47, 0.50, 0.55). 
(a) Mn $2p-3d$ resonance-maximum ($h\nu$ = 643 eV) and off-resonance 
($h\nu$ = 636 eV) photoemission spectra of 
Nd$_{0.6}$Sr$_{0.4}$MnO$_3$ measured at 20 K in the FM phase. 
The spectral intensity has been normalized to the photon flux. 
(b) High-resolution Mn $2p-3d$ resonance-maximum spectra 
representing the Mn $3d$ contributions in the PI (300 K for $x$ = 0.40 
and 0.47, 280 K for $x$ = 0.50 and 260 K for $x$ = 0.55), 
FM (20 and 170 K for $x$ = 0.40 and 0.47, 170 K for $x$ = 0.50) 
and COI (120 K for $x$ = 0.50) phases. 
The spectrum for $x$ = 0.55 measured in the antiferromagnetic metal 
(AFM, 20 K) phase is also added for comparison. 
(c) Same as (b), but focused on the $e_g$ spectral weight. 
(d) Detailed temperature dependence of the Mn $3d$ spectral weights 
near $E_F$ for $x$ = 0.40, 0.47 and 0.50 in the PI, FM and COI 
(120 K for $x$ = 0.50) phases. 
The spectral weights at different temperatures in (b)-(d) have been 
normalized to the integrated intensity from $E_F$ to $\sim$4 eV 
for each compound.}
\end{figure}

The photoemission data so far accumulated on 
manganites~\cite{TS95,JHP96,DDS96,AC97,JHPN98,AS99,Chuang01,JSK03} 
could not satisfactorily check the applicability of the theoretical models 
since most of the valence-band photoemission spectra have been 
obtained with low-energy light sources. 
The low-energy photoemission ($h\nu \lesssim$ 120 eV) spectra of 
three-dimensional transition-metal (TM) oxides mainly reflect 
the O $2p$ electronic states and moreover surface electronic structures, 
which might deviate strongly from the bulk TM $3d$ states, 
because of the relatively large ratio of the O $2p/$TM $3d$ 
photoionization cross sections~\cite{Lindau} and a short photoelectron 
mean free path~\cite{Tanuma} ($\lambda <$ 5 {\AA}, 
bulk contribution of $<30$ {\%}). 
The Mn $2p-3d$ resonance photoemission ($h\nu \sim$640 eV with 
$\lambda \sim$13 {\AA}, leading to a predominant bulk contribution 
of $\sim$60 {\%}) is promising for a more adequate discussion 
of the Mn $3d$ bulk electronic states and their temperature dependence 
since the Mn $3d$ spectral weight is selectively and drastically 
enhanced at the Mn $2p$ absorption threshold as shown in 
Fig.~\ref{Fig1}(a). 
The Mn $3d$ enhancement is much weaker and the bulk sensitivity 
is much less at the Mn $3p-3d$ resonance 
excitation, $h\nu \sim$50 eV.~\cite{TS95} 
However, it has been difficult for a long time to obtain the 
temperature-dependent Mn $3d$ spectra at the $2p-3d$ resonance 
with high resolution, 
because both photon flux of conventional synchrotron light sources 
and energy resolution of soft x-ray monochromators have been insufficient. 
It is inevitable to use a high-brilliance and high photon flux 
synchrotron light source in order to obtain the Mn $2p-3d$ resonance photoemission spectra with such a high energy resolution and 
good statistics as presented here. 
This has allowed us for the first time to observe the detailed 
temperature dependence of the electronic structures of manganites. 
In this Letter, we show a unified picture of the driving mechanisms 
leading to the various aspects of 
the $3d$ electronic states in the manganese perovskites. 

Here we report on the system Nd$_{1-x}$Sr$_x$MnO$_3$ with $x$ = 0.40, 
0.47, 0.50. 
The compounds with $x$ = 0.40 and 0.47 undergo a paramagnetic 
insulator (PI)-to-ferromagnetic metal (FM) transition at $T_c \sim$290 
and 275 K, respectively, where the FM state is stable down to the lowest temperatures below $T_c$.~\cite{Kajimoto} 
The PI-FM transition is also seen for $x$ = 0.50 ($T_c \sim$255 K), 
while this compound is a charge-ordered insulator (COI) below 
$T_{\rm COI} \sim$160 K.~\cite{KuwaSci} 
The photoemission measurements were performed at BL25SU in 
SPring-8.~\cite{RSI} 
The overall energy resolution was set to $\sim$100 meV. 
The base pressure was about 4 x 10$^{-8}$ Pa. 
For the temperature-dependent measurements, clean surfaces were 
obtained by fracturing the single crystalline samples {\it in situ} 
at 280-300 K in the PI phase. 
Then the spectra were measured at the resonance-maximum 
($h\nu$ = 643 eV) on cooling. 
A possible Auger contribution is clarified to be negligible 
at $h\nu$ = 643 eV within the energy region from $E_F$ to $\sim$3 eV. 
The surface cleanliness was confirmed before and after the measurements. 
We also checked that the Mn $3d$ spectra at low temperatures 
with fracture at low temperatures were very similar to these results.

The temperature dependence of the Mn $2p-3d$ on-resonance spectra 
(Mn $3d$ spectra hereafter) of Nd$_{1-x}$Sr$_x$MnO$_3$ is summarized 
in Figs.~\ref{Fig1}(b)-(d). 
The $t_{2g}$-derived electronic states appear as a strong peak 
near 2.0 eV. 
The spectral weight from $E_F$ to $\sim$1.5 eV originates dominantly 
from the $e_g$ states. 
These qualitative spectral features are in agreement with 
previous results~\cite{JHP96,AS99,JSK03} 
Surprisingly, however, the Mn $3d$ spectral line shapes change 
gradually with temperature not only in the $e_g$ region 
but also in the $t_{2g}$ region. 
Such spectral changes over a range of several eV within the same phase 
are very unusual. 
The $t_{2g}$ peak shifts gradually towards $E_F$ for $x$ = 0.40 and 
0.47 by about 0.15 eV between 300 and 20 K. 
On the contrary, the $t_{2g}$ as well as the $e_g$ components shift 
abruptly to higher binding energies concomitant with the opening of 
a clear energy gap of the order of 100 meV (Fig.~\ref{Fig1}(d)) 
for $x$ = 0.50 as a result of the FM-COI transition. 
As shown in Fig.~\ref{Fig1}(b), the gradual $t_{2g}$ peak shift 
for $x$ = 0.40 and 0.47 is not observed for 
Nd$_{0.45}$Sr$_{0.55}$MnO$_3$ ($x$ = 0.55), which undergoes a transition 
from the PI to antiferromagnetic metal (AFM) at 220 K.~\cite{KuwaPRL99} 
These results indicate that the gradual $t_{2g}$ peak shift towards 
$E_F$ with decreasing temperature for $x$ = 0.40 and 0.47 is 
characteristic of the PI-FM transition. 

A broad peak at 0.8-0.9 eV is observed in the PI phase 
as indicated by dashed bars in Fig.~\ref{Fig1}(c). 
We emphasize that this structure survives even in the FM phase 
at high temperatures while it is remarkably suppressed at 20 K 
for $x$ = 0.40 and 0.47 accompanied by a change of the spectral shape. 
Namely, another broad peak is seen at 0.5-0.6 eV as indicated 
by solid bars in Fig.~\ref{Fig1}(c). 
According to detailed measurements [see Fig.~\ref{Fig1}(d)], 
the intensity from $E_F$ to $\sim$0.3 eV is gradually enhanced 
with decreasing temperature in the FM phase. 
A clear Fermi cut-off is seen at low temperatures 
for $x$ = 0.40 and 0.47. 

To date, there has been a controversy whether the spectral function 
changes with temperature within the FM phase 
based on the low-energy photoemission data.~\cite{JHP96,DDS96,AC97} 
Our results clearly show that the Mn $3d$ spectral functions 
in both the $t_{2g}$ and the $e_g$ regions {\it do }change 
within the FM phase. 
Furthermore, there are two additional significant features in our spectra. 
One is that the spectra from 0.3 eV to $E_F$ remains essentially unchanged 
across the PI-FM transition [see Fig.~\ref{Fig1}(d)], 
in clear contrast to the remarkable change of the spectra 
across the FM-COI transition for $x$ = 0.50. 
The other is that the photoemission intensity at $E_F$ is distinctly 
finite even in the PI phase, 
which has never been seen before in the low-energy photoemission 
data.~\cite{JHP96,DDS96,AC97,JHPN98,Chuang01} 
We emphasize that the observed finite intensity at $E_F$ 
in the PI phase is intrinsic and cannot 
be reproduced by an instrumental resolution 
broadening of any plausible spectral line shape assumed to become zero 
intensity at $E_F$.~\cite{PESPI} 
The finite intensity at $E_F$ indicates that the compounds in the 
PI phase are neither simple band insulators nor Mott insulators as 
induced by an on-site Coulomb interaction between the Mn $3d$ electrons. 
It is thus found that the $3d$ electron correlation effects alone, 
which themselves are strong as previously pointed 
out,~\cite{TS95,DDS96,AS99} cannot account for the PI-FM transition 
in the manganites. 

We now turn to a discussion of the existing theoretical models against 
this anomalous temperature dependence of the Mn $3d$ spectral functions. 
The key issues of our data are summarized as follows: 
(a) finite intensity at $E_F$ even in the PI phase, 
(b) $t_{2g}$ peak shift with temperature in the FM phase, 
(c) presence of a peak at 0.8-0.9 eV in the PI phase, and 
(d) gradual intensity reduction of this peak and the appearance of 
another peak at 0.5-0.6 eV with decreasing temperature in the FM phase. 
Although the calculations based on the DE 
mechanism~\cite{Kubo72,Furukawa95} are not contradictory to (a), 
they cannot predict the $e_g$-derived broad peak away from $E_F$ at all, 
namely, fail to explain (c) and (d). 
The dynamical JT model~\cite{MillisPRL96,MillisPRB96} 
and a model including the orbital degrees of 
freedom~\cite{SYPRB98,SYPRL98} seem to fairly explain (a) and (c) 
since they show the broad-peak structure away from $E_F$. 
However, such a peak is hardly shifted with temperature 
according to these models, 
which is contradictory to (d).
In addition, these models are not likely to explain (b) because 
the center of gravity in the calculated $e_g$ spectral weights 
is hardly shifted with temperature. 
To conclude, all theoretical approaches mentioned above are 
still insufficient even for a qualitative explanation of 
the temperature dependence of the Mn $3d$ spectra. 
A further elaborated theoretical model is required 
to explain our experimental findings. 

\begin{figure}
\includegraphics[width=6cm,clip]{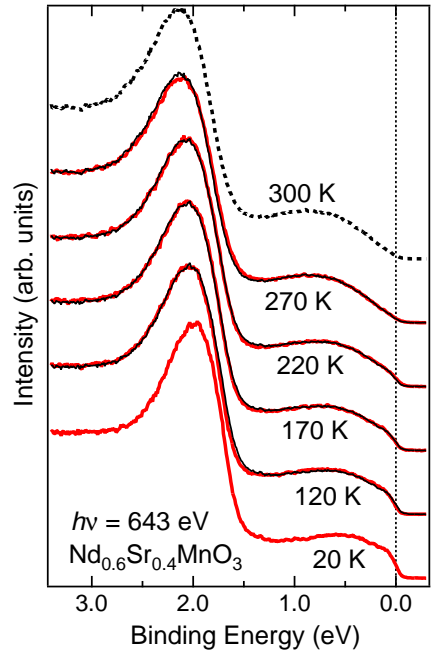}%
\caption{\label{Fig2}Detailed temperature dependence of the Mn 
$2p-3d$ resonance photoemission spectra of Nd$_{0.6}$Sr$_{0.4}$MnO$_3$. 
The red solid lines show the spectra in the FM phase 
while the black dashed line is in the PI phase at 300 K. 
Proper linear combinations of the two experimental spectra at 20 K 
and 300 K are shown by the black thin solid lines 
at intermediate temperatures, 
which can well reproduce the experimental results. 
In the linear combinations, the relative contributions of the 20 K 
spectrum are 68, 54, 43 and 7 \% at 120, 170, 220 and 270 K, 
respectively, with an accuracy of $\pm$7 \%. 
We have also successfully reproduced the temperature-dependent Mn $3d$ 
spectra of Nd$_{0.53}$Sr$_{0.47}$MnO$_3$ in the FM phase (not shown) 
by the linear combination of the spectra at 20 K and 300 K, 
where the contributions of the spectrum at 20 K are 50, 29 and 4 \% 
at 170, 220 and 270 K, respectively.}
\end{figure}

In order to understand the temperature-dependent spectral functions 
as much as possible from experimental viewpoint, 
we have tried to reproduce the spectra for $x$ = 0.40 and 0.47 
measured at 120, 170, 220, and 270 K in the FM phase 
by a linear combination of the spectra at 20 K in the FM phase 
and those at 300 K in the PI phase. 
As shown in Fig.~\ref{Fig2} for $x$ = 0.40, the linear combination of 
the spectra well reproduces the Mn $3d$ spectra in both the $e_g$ 
and the $t_{2g}$ regions at intermediate temperatures. 
This analysis indicates that the spectra in the FM phase still include 
the spectral component of the PI phase, 
whose weight is gradually reduced with decreasing temperature. 
Such a scenario that a dynamic and microscopic phase segregation 
grows in the FM region with increasing temperature, 
as has been proposed from inelastic neutron scattering on 
La$_{0.7}$Ca$_{0.3}$MnO$_3$,~\cite{Zhang01} is the most likely story 
to explain our data. 

In the following we show a unified picture, 
which explains the various electronic phases in the manganites. 
We focus now on the $e_g$ spectral weight and its temperature and 
doping dependence. 
The spectral shape in the PI phase is essentially similar for different 
$x$ as shown in Fig.~\ref{Fig1}(b) and \ref{Fig1}(c). 
In the ground states at the lowest temperature, on the other hand, 
the peak in the $e_g$ region becomes more prominent for $x$ = 0.50 (COI) 
and 0.55 (AFM) at higher binding energies compared with the PI phase, 
while it becomes broader for $x$ = 0.40 (FM) and 0.47 (FM) in the sense 
that the $e_g$ states extend above $E_F$ (remember that the clear Fermi 
cut-off is seen at low temperatures). 
This tendency corresponds well with the static JT distortion 
which is stronger for $x$ = 0.50 and 0.55 than for $x$ = 0.40 and 0.47 
at low temperatures.~\cite{KuwaSci,Kajimoto,KuwaPRL99} 
It is thought that the $e_g$ electron partially occupies the lower JT 
split band and hence provides more spectral weight 
at higher binding energies in the $e_g$ region. 
Accordingly the observed temperature dependence and the dynamic and 
microscopic phase segregation scenario are well understood as follows: 
The static JT distortion (i.e., the degree of freedom 
in the $e_g$ orbitals) mainly controls the $3d$ electronic states 
in the ground states at low temperatures. 
In particular, the completely developed FM phase without phase 
segregation can be realized at low temperatures for the compositions 
$x$ = 0.40 and 0.47 in which the static JT distortion is weak, 
while the FM ordering is suppressed for $x$ = 0.50 and 0.55 
where the JT distortion is strong. 
In the PI phase near room temperature, on the other hand, 
the dynamical JT effect governs the $3d$ electronic states irrespective 
of both $x$ and the strength of the static JT distortion 
at low temperatures. 
The dynamical JT effect weakens gradually with decreasing temperature 
below $T_c$. 
Then the amount of the FM phase grows correspondingly for 
0.40 $\leq x \leq$ 0.50. 
In this way, the dynamical and static JT effects hinder the FM ordering 
in the manganites at high and low temperatures, respectively. 

Finally we point out a close relation between the CMR and 
the dynamic phase segregation. 
The CMR effect is not notable near the phase boundary between the PI 
and FM states for Nd$_{1-x}$Sr$_x$MnO$_3$ ($x$ = 0.40-0.50) 
compared with that between the FM and COI states for 
Nd$_{0.5}$Sr$_{0.5}$MnO$_3$. 
From the above discussion, we conclude that the dynamic phase segregation 
originating from the gradually changed dynamical JT effect 
with temperature plays an important role in reducing the CMR 
at the PI-FM transition. 
Namely, in such a situation as the dynamical JT effect is strong 
in the PI phase and abruptly suppressed at the PI-FM transition, 
CMR would be remarkable. 

In conclusion, we have clearly observed the temperature dependence of 
the high-resolution Mn $2p-3d$ resonance photoemission spectra of 
Nd$_{1-x}$Sr$_x$MnO$_3$ in the PI and FM phases as well as 
evidence for the microscopic and dynamic phase segregation. 
It is clarified that Nd$_{1-x}$Sr$_x$MnO$_3$ in the PI phase are 
found to be neither a band insulator nor a Mott insulator. 
Although our results cannot be well explained by any available model, 
we propose a unified picture to interpret the experimental results. 
The $3d$ electronic states are mainly controlled by the static JT 
effect at low temperatures 
whereas the dynamical JT effect governs the electronic states 
near room temperature. 

We thank K. Noda, K. Konoike, T. Satonaka, S. Kasai, A. Yamasaki, 
A. Irizawa, M. Tsunekawa, and the staff of SPring-8, 
especially T. Muro, Y. Saitoh, and T. Matsushita 
for supporting the experiments. 
This work was supported by a Grant-in-Aid for COE Research from the 
Ministry of Education, Culture, Sports, Science and Technology (MEXT), 
Japan. 
AS acknowledges the support form the Kurata foundation. 
The photoemission measurements were performed under the approval 
of the Japan Synchrotron Radiation Research Institute 
(2001A0129-NS-np, 2003A0593-NS1-np).

\end{document}